# Estimands in Real-World Evidence Studies

Jie Chen[1], PhD, Daniel Scharfstein[2], ScD, Hongwei Wang[3], PhD,
Binbing Yu[4], PhD, Yang Song[5], PhD, Weili He[3], PhD, John Scott[6], PhD,
Xiwu Lin[7], PhD, Hana Lee[6*], PhD

[1]Biostatistics & Computing Science, Elixir Clinical Research, Warrington, PA

[2]University of Utah School of Medicine, Salt Lake City, UT

[3]Data and Statistical Sciences, AbbVie, North Chicago, IL

[4]Statistical Innovation, AstraZeneca, Gaithersburg, MD

[5]Biostatistics, Vertex Pharmaceuticals, Boston, MA

[6]FDA, Silver Spring, MD

[7]Janssen Pharmaceuticals, Horsham,  PA

*Correspondence to: Dr. Hana Lee, hana.lee@fda.hhs.gov, CDER, FDA, Silver Spring, MD 20903.

# Estimands in Real-World Evidence Studies

### Abstract

A Real-World Evidence (RWE) Scientific Working Group (SWG) of the American Statistical Association Biopharmaceutical Section (ASA BIOP) has been reviewing statistical considerations for the generation of RWE to support regulatory decision-making. As part of the effort, the working group is addressing estimands in RWE studies. Constructing the right estimand—the target of estimation—which reflects the research question and the study objective, is one of the key components in formulating a clinical study. ICH E9(R1) describes statistical principles for constructing estimands in clinical trials with a focus on five attributes—population, treatment, endpoints, intercurrent events, and population-level summary. However, defining estimands for clinical studies using real-world data (RWD), i.e., RWE studies, requires additional considerations due to, for example, heterogeneity of study population, complexity of treatment regimes, different types and patterns of intercurrent events, and complexities in choosing study endpoints. This paper reviews the essential components of estimands and causal inference framework, discusses considerations in constructing estimands for RWE studies, highlights similarities and differences in traditional clinical trial and RWE study estimands, and provides a roadmap for choosing appropriate estimands for RWE studies.

## 1   Introduction

One of the most essential steps in clinical research is to clearly state the research question which determines the estimand—a quantity that connects the research objective to the target of inference. An estimand goes beyond statistical procedures; indeed, it drives all subsequent steps including design, conduct, and analysis of study, as well as interpretation of results. The International Council for Harmonisation of Technical Requirements for Pharmaceuticals for Human Use (ICH) E9(R1) guidance (ICH, 2021) presents principles for constructing estimands in clinical trials with a focus on five attributes—population, treatment, endpoints, intercurrent events, and population-level summary. In particular, the guidance elucidates the importance of handling intercurrent events (e.g., discontinuation of assigned treatment, use of alternative/additional medications, etc.) in defining an estimand. Further discussions on estimands in clinical trial settings can be found in Akacha et al.

(2017), Mallinckrodt et al. (2019), Lipkovich et al. (2020), Ratitch et al. (2020), Qu et al. (2020), Qu et al. (2021), Hernán and Scharfstein (2018), Scharfstein (2019) and references therein.

While some researchers differentiate clinical trials and real-world studies based on the use of randomization (Collins et al., 2020), the U.S. Food and Drug Administration (FDA) does not. According to the Framework for FDA's Real-World Evidence Program (FDA, 2018), RWE is defined as "the clinical evidence about the usage and potential benefits or risks of a medical product derived from analysis of real-world data", and thus randomized controlled trials (RCTs) can also generate RWE depending on their use of RWD. The FDA's RWE Framework provides various examples of RCTs using RWD to generate RWE including a pragmatic trial and a registry-based multi-center trial. We adapt the FDA's perspective on RWE and consider real-world estimands to be *target quantities for clinical studies for medical product development which intend to generate evidence from analysis of RWD.* In this article, we will refer to such clinical studies, which encompass both RCTs and non-randomized studies, as *real-world evidence studies* (RWE studies). However, this article will differentiate RWE studies from traditional clinical trials (TCTs), following the FDA's RWE framework which states that "evidence from traditional clinical trials will not be considered RWE". TCTs in this paper refer to RCTs that (a) do not involve analysis of RWD for decision making, (b) are designed to control variability and to maximize data quality, and (c) are supported by a research infrastructure that is largely separated from routine clinical practice. With this in mind, the ASA BIOP RWE SWG has spent a significant amount of effort to clarify estimands in RWE studies, which is summarized in this article. Specifically, we intend to (1) elucidate the similarities and differences of estimands between TCTs and RWE studies, (2) present points-to-consider when defining an estimand, (3) discuss illustrative examples of estimands, and (4) provide a roadmap for choosing an appropriate estimand in RWE studies.

The rest of the paper is organized as follows. Section 2 reviews some useful frameworks for defining estimands. Section 3 discusses considerations in defining estimands for RWE studies by focusing on the five attributes of estimands in ICH E9(R1) (ICH, 2021). This section also provides a summary of similarities and differences between TCT estimands and RWE study estimands. Section 4 delineates specific examples of estimands for TCTs and various types of RWE studies, and Section 5 depicts a roadmap for choosing appropriate estimands in RWE studies. Section 6 provides a discussion and concluding remark.

## 2 Frameworks for Defining Estimands

The ICH E8(R1) Guideline on General Considerations for Clinical Trials (ICH, 2019) stresses that clinical studies intended to support approval of medical products should be based on explicitly stated study objectives that address specific scientific questions. The ICH E9(R1) addendum on estimands and sensitivity analysis in clinical trials (ICH, 2021, henceforth E9R(1)) emphasizes the need for systematic approaches to translate the key clinical question and corresponding trial objective into an "estimand" to ensure alignment among the study objective, trial design, data collection, analysis, and interpretation. The key question is then: *How can we formally define the estimand that best describes the study objective*? Different frameworks have been suggested to construct an estimand. These frameworks are distinguished by different specifications of the essential components (attributes) of an estimand. In this section, we review some of these frameworks and discuss the attributes within each framework that are most relevant to defining an estimand in RWE studies.

### 2.1 ICH E9(R1) Statistical Principles for Clinical Trials: Addendum: Estimands and Sensitivity Analysis in Clinical Trials

E9(R1) (ICH, 2021) is intended to serve as a basis to strengthen the dialogue between regulatory agencies and sponsors about what a clinical trial should aim to demonstrate, as well as to facilitate communication across different stakeholders (e.g., patients, physicians, and sponsors, etc.) involved in trial formulations at various stages. The estimand framework in E9(R1) provides a basis for constructing an estimand not only for TCTs but also for RWE studies.

E9(R1) describes five attributes of an estimand and discusses considerations for its construction. For more details on each attribute, see section III.B (A.3.2) and III.C (A.3.3) of ICH (2021).

1. *Population.* Patients targeted by the clinical question. This attribute might include

the entire trial population or a subgroup of patients with particular characteristics.

2. *Treatment.* The treatment condition of primary interest (e.g., a new drug) and, as appropriate, the alternative treatment condition to which comparison will be made (i.e., comparator). The treatment might be an individual intervention, a combination of concurrent interventions (e.g., add-on therapy to standard of care (SOC)), or a complex sequence of interventions.

3. *Variable* (or endpoint). The specification of an endpoint might include whether a patient experiences an intercurrent event (e.g., composite outcome). The endpoint could be measured repeatedly during the course of treatment and/or up to a certain time point of interest, and thus often involves time component.

4. *Intercurrent events* (ICEs). Events occurring after treatment initiation that affect either (1) the interpretation or (2) the existence of measurements of endpoints associated with the clinical question of interest. Some examples of ICEs are:

   - Events affecting the interpretation of endpoints: e.g., treatment discontinuation or use of an additional/alternative therapy.
   - Events affecting the existence of the measurements: e.g., terminal events such as death, surgery, etc., when these events are not part of the endpoint itself.
   - Other types: Events (either occurrence or non-occurrence) that can be used to define principal strata, e.g., tumor shrinkage in oncology, infection, or severity of infection after vaccination in vaccine trials.

Note that ICEs in TCTs should be distinguished from missing data, though the distinction becomes complicated in RWE studies (see Section 3.4). E9(R1) states that "neither study withdrawal nor other reasons for missing data (e.g., administrative censoring in trials with survival endpoints) are in themselves intercurrent events." As an example, E9(R1) illustrates that "subjects who withdraw from the trial may have experienced an intercurrent event before the study withdrawal."

5. *Population-level summary.* A summary measure for the endpoint that provides a basis for comparison between treatment conditions. This is an effect measure such as risk difference, risk ratio, odds ratio, hazard ratio (HR), (restricted) mean survival time difference, etc.

Of note, these estimand attributes contain the key components of PICOT—Population, intervention, comparator, outcome, and time—which is often used to guide evidence generation in clinical research (Rios et al., 2010; Riva et al., 2012).

E9(R1) emphasizes pre-specification of plausible ICEs in constructing an estimand; see Section III.D (A.3.4) of the addendum regarding considerations on selecting an appropriate strategy to address potential ICEs and to construct an appropriate estimand. In brief, E9(R1) discusses five potential strategies—treatment policy, hypothetical, composite variable, while-on-treatment, and principal stratum. Strategies for handling ICEs may also be reflected in defining the target population (e.g., a principal stratum defined by the occurrence of a specific ICE) and treatment (e.g., discontinuation of assigned treatment due to intolerability may be considered as a failure if the outcome variable is success or failure).

Hernán and Scharfstein (2018) and Scharfstein (2019) suggest that trials should be designed based on clinically relevant treatment strategies, rather than assuming patients are expected to take a fixed dose of the treatment to which they are assigned. For example, they argue that an estimand which implicitly compares "do not treat" versus "treat even if contraindications arise" is irrelevant to both regulators and clinicians in practice. They further point out that from trial design and interpretation perspective, the focus should be on "estimands that compare clinically relevant and potentially implementable sustained treatment strategies in groups of patients defined by measurable characteristics." Scharfstein (2019) demonstrates that precise definition of the treatment strategy can lead to precise definition of an outcome without involving the notion of "intercurrent events" or "strategies to handle intercurrent events". Scharfstein (2019) proposes four attributes for an estimand: (1) target population, (2) treatment strategy, (3) outcome (intention-to-treat, composite, counterfactual, while adherent), and (4) effect measure. For a well-defined treatment strategy that specifies what a patient should do in the presence of, say, side effects or lack of efficacy (LoE), the outcome under full adherence (possibly counterfactual for non-adherent patients) to the treatment strategy is clinically relevant.

## 2.2   Causal inference framework

Clinical studies for medical product development are designed to estimate the casual effect of treatment, as the following statement on treatment effect in E9(R1) alludes to the notion of potential outcome (counterfactual): "*central questions for drug development and licensing are to establish the existence, and to estimate the magnitude, of treatment effects: how the outcome of treatment compares to what would have happened to the same subjects under alternative treatment (i.e. had they not received the treatment, or had they received a different treatment).*" Ho et al. (2023) and Lipkovich et al. (2020) demonstrate how the use of the causal inference framework and potential outcome language can provide a unified, systematic approach to define causal estimands for both randomized and non-randomized studies.

Consider a trial in which participants are randomized to one of two treatments and there may be immediate crossover. Let $Y(a, t)$ be the potential outcome under assignment to treatment $a$ and actual receipt of treatment $t$. Let $T(a)$ be receipt of treatment under assignment to treatment $a$ and $Y(a) = Y(a, T(a))$ be the outcome under assignment to treatment $a$. Note that $Y(a) = Y(a, T(a))$ is sometimes referred to as the composition assumption (Pearl, 2009), i.e., the potential outcome $Y(a)$ intervening to set $A$ to $a$ and to set $T$ to the value it would have been if $A$ had been $a$ (VanderWeele and Vansteelandt, 2009). With this notation, one can define a number of causal estimands. The intention-to-treat estimand is a contrast between the mean of $Y(1)$ and the mean of $Y(0)$, i.e., the difference in mean of the outcome in a world in which everyone is assigned to treatment 1 versus a world in which everyone is assigned to treatment 0. If we are willing to impose the exclusion restriction, i.e., $Y(0, t) = Y(1, t) = Z(t)$, then we may be interested in a hypothetical estimand that contrasts the mean of $Z(1)$ and the mean of $Z(0)$, i.e., the difference in mean of the outcome in a world in which everyone is forced to take treatment 1 versus a world in which everyone is forced to take treatment 0. Alternatively, there may be interest in a principal stratum estimand which contrasts the mean of $Y(1)$ and the mean of $Y(0)$ among "compliers" (i.e., $T(1) = 1$ and $T(0) = 0$).

In a randomized trial, suppose that the assigned treatment is to be taken at $K$ time-points. Now, let $\mathbf{a} = (a_1, \ldots, a_K)$ be a specified treatment sequence, where $a_k \in \{0, 1\}$ for $k = 1, \ldots, K$. Let $\mathbf{A} = (A_1, \ldots, A_K)$ be the actual treatment sequence received, where $A_k \in$ $\{0, 1\}$ for $k = 1, \ldots, K$. Define the potential outcome at time $K$ as $Y_K(\mathbf{a}) = Y_K(a_1, \ldots, a_K)$. An example of a hypothetical estimand is a contrast between the mean of $Y_K(\mathbf{1})$ and the mean of $Y_K(\mathbf{0})$, i.e., the difference in mean of the outcome in a world in which everyone takes treatment 1 at all $K$ timepoints versus a world in which everyone takes treatment 0 at all $K$ timepoints. Note that this is different from an intention-to-treat estimand.

In addition to time $K$, suppose that the outcome is measured at times $1, \ldots, K - 1$. Let $Y_k(\mathbf{a})$ be the potential outcome at time $k$ under treatment sequence $\mathbf{a}$. We assume that treatments after time $k$ do not influence $Y_k(\mathbf{a})$, i.e., $Y_k(\mathbf{a}) = Y_k(a_1, \ldots, a_k)$. Let $A^*$ be the randomized treatment. Suppose that adherence is monotone, i.e., if $A_k \neq A^*$ then $A_{k+1} \neq A^*$. Let $k^* = \max\{k : A_k = A^*, k = 1, \ldots, K\}$. For each participant, let $W$ be the average outcome while adherent to the assigned treatment, i.e., $W = \frac{1}{k^*} \sum_{k=1}^{k^*} Y_k(A^*, \ldots, A^*)$. The while-on-treatment estimand is a contrast between the mean of $W$ among patients assigned to treatment 1 and those assigned to treatment 0.

A general discussion of potential language for defining estimands in RWE settings including longitudinal and dynamic treatment regime studies can be found in Gruber et al. (2022) and Wu et al. (2023). To draw inference about these estimands, assumptions are required; the reasonableness of the assumptions will depend on the clinical setting. The potential outcome language is a precise and transparent way of defining estimands. An appealing feature of the causal framework is that it provides a quantitative formalization of estimands and assists with understanding the assumptions needed to estimate them from the available data. A limitation of this framework is that non-quantitative stakeholders are not as familiar with this language.

## 2.3   Target trial

In designing and analyzing an observational study to answer a specific causal research question, Hernán and Robins (2016) recommend target trial emulation. Key components of their framework, which are called the *target trial protocol components*, include specification of (1) eligibility criteria, (2) treatment strategies, (3) treatment assignment mechanism, (4) time zero (baseline) and follow-up period, (5) outcome, (6) causal contrasts of interest, and (7) analytic strategy. The analytic strategy lays out the assumptions (some untestable) and inferential procedures to learn about the specified causal contrasts of interest. These assumptions will help clarify key data collection elements (e.g., baseline and post-baseline covariates). There are close connections between the target trial protocol components and the attributes of an estimand discussed in E9(R1), Hernán and Scharfstein (2018), and Scharfstein (2019). For example, protocol components (1) and (5) in the target trial correspond to population and endpoint attributes in E9(R1), respectively. Protocol component (2) reflects treatment (and may also include some ICEs), and protocol component (6) aligns with population-level summary in E9(R1).

Given its focus on observational studies, the target trial framework is useful for constructing estimands in real-world settings. For example, the target trial dictates importance of emulating baseline randomization using pre-exposure confounding variables and considering the issue of time zero which were not discussed in E9(R1) as its primary focus in defining estimand for TCT settings. In contrast, E9(R1) discusses ICEs, which are not explicitly discussed in the target trial framework. However, this can be handled through proper outcome definition.

Hernán and Robins (2016) point out that the target trial emulated by observational databases will typically be a pragmatic trial as it is impossible to emulate a TCT involving scheduled outcome assessments and monitoring/encouragement of adherence to treatment regimes. Importantly, failure to emulate a reasonably defined target trial should serve as a red flag as to the quality of the scientific investigation. Here, "a reasonably defined target trial" means that it might be less stringent than the most ideal trial setting due to inherent limitations in RWD source, design components, etc. For example, RWE studies utilizing Medicare claims data are ideal for capturing drug exposure but may be limited in identifying cancer diagnoses and potentially important covariates such as smoking. This can be mitigated by using validated algorithms for cancer identification with the use of International Classification of Diseases (ICD), National Drug Code (NDC), and/or Healthcare Common Procedure Coding System (HCPCS) to capture evidence of disease diagnoses, smoking, and smoking status in the claims data. Such RWE studies might still not be a perfect emulation of their target trial, but could be considered an emulation of "a reasonably" defined target trial.

We have reviewed three frameworks that can help define estimands in clinical trials. In the rest of the paper, we will expand our discussion on estimand for RWE studies by focusing on the five attributes of estimands in ICH E9(R1) as they help precisely construct causal estimands of interest in clinical studies using non-technical language and hence facilitating communication among various disciplines involved in the formulation of RWE studies.

## 3  Estimands in RWE Studies

This section describes points-to-consider when defining an estimand in RWE studies with a focus on the five attributes in E9(R1). It also provides some discussion on assessment of fit-for-purpose RWD and covariates that are important in creating comparable groups based on RWD. Relatively speaking, covariate adjustment is less imperative for randomized studies with good treatment adherence and follow-up, but is a major consideration in RWE studies.

### 3.1  Population—heterogeneity, eligibility, and representation

The target population for RWE studies is typically more heterogeneous than that of TCTs. For example, patients having comorbidities and using concomitant medications may be eligible for RWE studies as they might represent the target population under routine clinical care practice. RWE studies may include patients with more diverse demographic backgrounds representing various geographical locations and different healthcare systems (e.g., Medicaid/Medicare, commercial insurance programs, academic medical centers, community-based clinics, etc.) Therefore, population heterogeneity in RWE studies may capture and reflect (i) patients who are eligible for but choose not to participate in a TCT or (ii) under-represented patient populations (e.g., ethnic minorities, elderly, and those residing in remote locations) who are eligible for, but otherwise do not have access to, facilities of a TCT (Unger et al., 2019).

On the other hand, some RWD may be less representative of the general population. For example, commercial insurance claims data usually cover those who are medically insured through employment and thus relatively healthier than general population, as evidenced by lower mortality and morbidity rates (i.e., inducing the so-called healthy worker effect (Li and Sung, 1999)). In addition, some community-based RWD can be highly restrictive, such as the Framingham Heart Study which originally included individuals of predominantly white European descent, although some minority residents of Framingham were recruited later (Mahmood et al., 2014).

## 3.2 Treatment—adherence, preferences, and dynamic treatment regimes

Patients in an RWE study may exhibit more complex treatment patterns relative to those in a TCT (Ogundipe et al., 2021; Nicholas et al., 2020). For example, patients with type 2 diabetes who are initially prescribed dipeptidyl peptidase-4 (DPP4) inhibitors generally have low adherence and exhibit intra-class switches (Ogundipe et al., 2021). In addition, patients and prescribers may react to LoE or adverse events differently, leading to treatment non-adherence, dosage adjustment, treatment switching, concomitant use of another medication, or initiation of some dynamic treatment regime that adjusts treatment based on accumulated patient information (Hernán and Scharfstein, 2018; Scharfstein, 2019; Li et al., 2023). All of these reflect "real" clinical practice, and thus should be considered as part of a treatment (regime) defining attribute and should be reflected in estimand definition in RWE studies.

Because of the complexity of treatments in RWE studies, it is important to clearly articulate the treatment strategy/regime of primary interest in the study protocol. In addition, adherence to the selected treatment strategy should be closely monitored, as suggested in Scharfstein (2019). From a statistical perspective, it is helpful to use the potential outcome framework to precisely define the estimand of interest.

## 3.3 Endpoints—Surrogate endpoints, clinical outcomes, and survival

An endpoint is "a precisely defined variable intended to reflect an outcome of interest that is statistically analyzed to address a particular research question" (FDA, 2016). In RWE studies, the choice of endpoint(s) depends not only on the research question, but also on the available RWD sources. The research question is based on the concept of interest, defined as "the aspect of an individual's clinical, biological, physical, or functional state, or experience that the outcome assessment is intended to capture or reflect" (FDA, 2016). However, RWD resources determine whether an endpoint can be reliably and validly captured (Levenson et al., 2023).

Endpoints can be divided into two broad categories: (1) surrogate endpoints (or biomark-ers) measuring biologic processes, pathogenic processes, or responses to an exposure or intervention and (2) clinical endpoints (or outcomes) measuring how a patient feels, functions, or survives (FDA, 2016). Unlike in TCTs, surrogate endpoints are less likely to be used as endpoints in RWE studies. Surrogate endpoints may not be assessed (and hence the values are not collected) or they may require validated biotechnologies (e.g., bioassays) that are not universally available across all sites from which the RWD are collected (Mercon et al., 2020). Also note that simple clinical outcomes reflecting only one specific condition such as death or hospitalization may be more appropriate when composite outcomes containing multiple components in RWE studies. This is because some components of a composite outcome (e.g., eczema area severity index) may not be captured in RWD or may involve subjective measures that require additional effort for standardization which may not be possible (FDA, 2021).

The use of clinical outcomes as endpoint variables in RWE studies needs careful consideration. First, not all relevant clinical outcomes may be recorded; patients may not be assessed if they relocate, die, or experience improved conditions. Whether an outcome is recorded may also depend on an individual physician's discretion (e.g., clinician reported outcomes). Second, outcomes may be reported at irregular, possibly spontaneous times whenever a patient visits his/her physician. Third, outcomes may be over- or under-reported if they are subjective; a patient-reported outcome (PRO) may or may not be captured depending on how the patient feels or functions. Fourth, clinical outcomes that are not adjudicated by an independent party may be subject to information bias. Lastly, outcomes may not be directly captured but need to be derived from various RWD sources; see the ASA BIOP RWE SWG paper by Levenson et al. (2022) on statistical consideration for assessment of fit-for-purpose real-world data to support regulatory decision-making in drug development.

### 3.4 Intercurrent events—drug-induced, behavioral and non-behavioral adherence

E9(R1) (ICH, 2021) highlights the importance of ICEs when defining an estimand. Some commonly encountered ICEs include non-adherence to assigned treatment due to adverse events (AE) or LoE, adjustment of assigned dosage level, use of rescue medications, and terminal events such as death. Qu et al. (2021) classify ICEs into three broad categories: (1)

Category 1 ICEs potentially associated with safety, which explains the reasons for patients who are unable to tolerate and hence discontinue their assigned treatment, (2) Category 2 ICEs potentially due to LoE which may include the use of rescue medication, switching to a higher dose of the assigned therapy, or to a new therapy, and (3) Category 3 ICEs due to all the other reasons (e.g., death, loss-to-follow-up). This categorization is useful when defining an estimand for TCTs where the choice of treatments and the occurrence of ICEs are less influenced by the patient-physician preference. However, treatment regimes and types of ICEs (and their occurrence) in RWE studies are much more complex as described in Section 3.2. For more informed decision-making, the clinical questions of interest and corresponding treatment strategy/regime should reflect the patient's perspective (including their preferences, values, behaviors, and lifestyle) as well as features of routine medical practice. Towards this, one may consider the following five categories of ICEs for RWE studies:

E1 *Events due to safety concerns*: This category is coincident with Category 1 in Qu et al. (2021).

E2 *Events due to lack of efficacy*: This category is coincident with Category 2 in Qu et al. (2021).

E3 *Events related to behavioral factors*: ICEs potentially associated with patient behavior-related factors such as (1) patient preference, (2) inconvenience of use for the current product, (3) recommendations by a friend or family member to switch to another therapy, and (4) physician-patient relationship.

E4 *Events related to non-behavioral factors*: ICEs potentially associated with non-behavior-related factors such as (1) change of health insurance program (e.g., the new program doesn't cover the current therapy), (2) relocation to another place where the current therapy is unavailable, (3) development of new conditions that contradict the use of the current therapy, (4) improvement of health conditions, and (5) participation in a clinical trial that requires discontinuation of the current therapy.

E5 *Terminal events* (TEs): TEs are events that affect the existence and/or measurements of endpoints while not being part of endpoint defining attributes. For example, diabetic foot ulcers cannot be assessed if a diabetic patient died or experienced leg amputation.

Categories E1 and E2 are product-induced ICEs and may be captured in RWD. However, some under-reporting can be expected, e.g., LoE or reasons for switching therapy might not be recorded. The distinction between E3 and E4 events can be helpful in some medical research settings. For example, certain patient behavioral characteristics might be major determinants of treatment discontinuation, regardless of the medical product effect itself. If this is the case, differentiating E1 to E4 ICEs might be critical in defining the estimand; the intention-to-treat estimand would be a combined effect of the medical product and patient's behavior/lifestyle. One might be tempted to define an estimand involving a hypothetical world in which discontinuation of treatment due to patient behavior is ruled out. Category E5 is a set of special events that may be treated differently depending on the research question. In some situations, it may make sense to incorporate death into the definition of a composite endpoint (Kahan et al., 2020) as, for example, progression-free survival in oncology studies.

As mentioned earlier, RWD may not capture information on adherence. Multiple ICEs may occur on the same patient at different time points and the chronological order of occurrence may be unknown. Some ICEs (e.g., treatment switch, use of rescue medications) might be a part of (dynamic) treatment strategies. Therefore, we emphasize that a RWE study protocol should consider as many ICEs as possible, prioritize them if there are multiple ICEs, and clearly articulate the primary treatment (strategy) of interest and differentiate them from secondary treatment (strategies) of interest. Note that ICEs should be clinically meaningful and plausible. In case of multiple ICEs, prioritization could be based on clinical impact, interpretability of estimand, and frequency of occurrence. The causal inference frameworks may be helpful to statisticians in articulating the role of ICEs when precisely defining the treatment regime of interest. In addition, E9(R1) and target trial frameworks can further support communication between statisticians and domain experts. Once the primary treatment regime is selected, RWD quality and RWE study design and conduct must ensure that the proposed RWE studies can capture adherence to the selected treatment strategy.

## 3.5 Population-level summary

In both TCTs and RWE studies, the choice of population summary will generally depend on the type of endpoint and the clinical research question. Statistical properties as well as the ease of interpretation should also be considered. For example, if the clinical question is about treatment effect on cancer patients as measured by the number of patients who respond to the treatment, then the difference, rather than the ratio, in response rate between groups could be chosen as a population-level summary because the former is always well-defined with zero count in the group at denominator (Miettinen and Nurminen, 1985); if the endpoint is progression-free survival time, then the difference in (restricted) mean survival time might be preferred over hazard ratio in a Cox regression model since the latter may not bear causal interpretation (Hernán, 2010; Aalen et al., 2015).

## 3.6 Sensitivity analysis

The identification (i.e., precisely known if the population were infinite) of all estimands from the observed data relies on assumption(s). A clear benefit of RCTs is that the treatment assignment mechanism is via an external random process. This greatly facilitates inference about certain estimands, such as one based on intention-to-treat, as the assumptions required for identification hold by design. Other estimands in RCTs and all estimands in non-randomized RWE studies, will require untestable assumptions, e.g., no unmeasured confounders. If these assumptions fail to hold, then inference may be biased. As a result, it is essential to conduct sensitivity analysis to evaluate the robustness of inferences. Scharfstein (2019) suggests rigorous sensitivity analysis to understand whether "there are reasonable assumptions with scientific justification under which the inferences about the estimand can be suggestive of no clinical benefit or even harm." See Zhang et al. (2018) for detailed review of methodologies for sensitivity analysis to evaluate the impact of unmeasured confounding in observational studies.

## 3.7 Other considerations

Construction of an estimand should consider not only what is of clinical relevance for the treatment of interest in a specific therapeutic setting, but also whether an estimate of the treatment effect can be reliably derived for decision-making. The E9(R1) points out that "an iterative process should be used to reach an estimand that is of clinical relevance for decision-making, and for which a reliable estimate can be made." (ICH, 2021). This is particularly important in defining an appropriate estimand for RWE studies. We provide a brief discussion below on additional considerations from the perspectives of fit-for-use RWD and covariates.

*Fit-for-use RWD.* The framework for FDA's RWE program (FDA, 2018) defines fit-for-use RWD in terms of reliability and relevance. Reliability concerns data accrual and data quality control, while relevance considers whether captured data contains essential data elements such as exposure, outcomes, and covariates, with the latter being critical for defining an appropriate estimand. Levenson et al. (2022) propose three-dimensional criteria for assessing fit-for-use RWD—Relevance, reliability, and fit-for-research, in which there are six elements in the relevance dimension based on a specific research question: disease population, outcome, treatment, confounders, time, and generalizability. While the first three elements (plus time) are already described as basic attributes of an estimand, confounders are important in non-randomized studies to allow adjustment for selection bias. Confounders are key covariates which are often used for estimand identification.

*Covariates.* Covariates are not among the five attributes of estimands in E9(R1) (ICH, 2021). However, the assumptions needed to identify certain estimands in RCTs and all estimands in non-randomized RWE studies rely critically upon them. A typical benchmark assumption involves some form of conditional independence, where the conditioning is based on a set of measured covariates. When the estimand involves a point treatment, the covariates are typically measured at baseline; when the estimand involves a time-varying treatment, the covariates may also be time-varying.

In RCTs, baseline covariates are typically measured just prior to treatment initiation, i.e., at the time of enrollment or randomization; therefore, the covariates are not impacted by treatment. In RWE studies, baseline covariates are often measured during a baseline *window or period* (e.g., 12 months prior to index date when treatment starts), which can lead to ascertainment bias (Anes et al., 2021). For example, one patient might have information on history of prior medication use for the full 12-month baseline window whereas another patient might only have information during the 3 months prior to the index date. Second, missing information on some covariates can lead to selection bias. For instance, laboratory test results may not be captured for all patients in the target population. Third, the reason for missingness of key covariates are usually unknown or not well-captured in RWD. Importantly, it is complicated to differentiate missingness (e.g., information intended to be collected but not recorded) from non-existence (e.g., information not intended to be collected as informed presence or absence) in RWD and thus it is hard to address the issue analytically. See the ASA BIOP RWE SWG paper by Levenson et al. (2022) for detailed discussion on informed presence or non-presence of missing values.

## 3.8 Similarities and differences in defining an estimand between TCTs and RWE studies

Given the above discussion, similarities and differences in defining estimands between TCTs and RWE studies with respect to the five attributes in E9(R1) are summarized in Table 1.

# 4 Examples of Estimands in TCT and RWE Studies

This section presents examples of estimands and their attributes in a TCT and a wide range of RWE studies including randomized and non-randomized studies.

## 4.1 Traditional clinical trials

The CAROLINA trial was a randomized, double-blind, active-controlled, non-inferiority trial, with participant screening from November 2010 to December 2012, conducted at 607 hospitals and primary care sites in 43 countries involving 6042 participants (Rosenstock et al., 2019). The trial was designed to address the following clinical question: *What is the effect of linagliptin compared with glimepiride on major cardiovascular events in patients with relatively early type 2 diabetes and elevated cardiovascular risk*? The study question was described by the following estimand attributes:

- *Population*: adult patients with type 2 diabetes, glycated hemoglobin of 6.5% to 8.5%, and elevated cardiovascular risk, plus a list of inclusion and exclusion criteria

- *Treatment*: at least 1 dose of study medication, either 5 mg of linagliptin once daily (n = 3023) where control is 1 to 4 mg of glimepiride once daily (n = 3010), both in addition to usual care

- *Endpoint* : time to first occurrence of cardiovascular death, nonfatal myocardial infarction, or nonfatal stroke (i.e., composite)

- *ICEs*: prematurely discontinuing the study drug (38.2% of participants), intensifying glycemic treatments (primarily by adding or adjusting metformin, glucosidase inhibitors, thiazolidinediones, or insulin, according to clinical need)

- *Population-level summary* : hazard ratio

Of note, there are two problems with this definition that pose analytic challenges. First, they have not addressed how death from another cause will be handled; censoring death from other causes is problematic. Second, requiring at least one dose of the study drugs can introduce bias as taking the drug is a post-randomization factor. To answer the clinical question of the study, the while-on-treatment strategy might be appropriate to restrict the observational time to the time period before the occurrence of ICEs. Note that care should be taken if the occurrence of ICEs differs between treatment and control arms.

## 4.2  Pragmatic trials

Vestbo et al. (2016) conducted a pragmatic trial to evaluate the effect of once-daily inhaled combination of fluticasone furoate at a dose of 100 ug and vilanterol at a dose of 25 ug among patients with chronic obstructive pulmonary disease (COPD). Patients were randomly assigned to either the fluticasone furoate-vilanterol group or the usual care group. The primary outcome was the rate of moderate or severe COPD exacerbations among patients who had an exacerbation within 1 year before the initial randomization. A key pragmatic feature of the trial is that it did not require regular, enforced follow-up assessments; outcomes were captured from the Electronic Medical Record (EMR). The study question was described by the following estimand  attributes:

- *Population*: patients who were 40 years of age or older, had received a documented diagnosis of COPD from a general practitioner, and had had one or more COPD exacerbations in the previous 3 years before enrollment

- *Treatment* : combination therapy with 100 ug of fluticasone furoate and 25 ug of vilanterol (the fluticasone furoate-vilanterol group), or the continuation of usual care as determined by the general practitioner (the usual-care  group)

- *Endpoint*: the number, within one year of randomization, of moderate or severe exacerbations, defined as any worsening of respiratory symptoms that led to treatment with antibiotic agents or systemic glucocorticoids (or both), to hospital admission, or to scheduled or unscheduled hospital visits

- *ICEs*: treatment switch from fluticasone furoate-vilanterol to usual care (patients in the usual-care group were not permitted to switch to the fluticasone furoate-vilanterol group), serious adverse events that cause discontinuation of any assigned treatment

- *Population-level summary*: the mean annual rate of moderate or severe exacerbations

Patients who never took fluticasone furoate-vilanterol were excluded. This might be problematic as taking the drug is a post-randomization factor and exclusion is restricted in only one of the treatment groups. The study uses a treatment policy (intention-to-treat) strategy in the analysis, which makes the use of randomization and ignores the occurrence of ICEs. This strategy might be appropriate if the switch from fluticasone furoate-vilanterol to the usual care reflects the clinical practice; however, the treatment effect could be estimated toward the null.

### 4.3   Single-arm trials

Single arm trials are commonly used to address clinical questions dealing with highly unmet medical needs with or without explicitly defining an external control group.

*Single-arm trials without explicit external controls.* This type of trial often aims to confirm the existence of a minimal, clinically meaningful treatment effect (often provided by historical experience with the best possible SoC or clinical guidance). An example is the TOCIVID-19 trial that evaluates efficacy and tolerability of tocilizumab in the treatment of 330 patients with severe or critical COVID-19 pneumonia (Chiodini et al., 2020). This single-arm study was designed to evaluate whether the 1-month mortality rate was less than 15%. The five attributes for the primary estimand are:

- *Population*: patients with severe or critical COVID-19 pneumonia

- *Treatment* : tocilizumab, regardless of dose and actual route, time, and number of administrations

- *Endpoint*: death/alive status 30 days after registration

- *ICEs*: eight pre-defined ICEs including delay in administering the treatment, discharge before day 14 or 30, and discharge or death before treatment

- *Population-level summary*: 1-month mortality rate

A treatment policy strategy was used to calculate the mortality rate. The study also defined time-to-death as a secondary endpoint and considered an estimand based on hypothetical world without "delay in administering the treatment due to shortage or administrative reasons" (Chiodini et al., 2020).

*Single-arm trials with external controls.* Some single-arm trials use RWD to construct an external control group, either historical or concurrent; see the FDA draft Guidance on Rare Diseases: Common Issues in Drug Development (FDA, 2019) and the ICH E10 (ICH, 2001) on the choice of control group in clinical trials for situations where external controls can be used. An example of single-arm trial using RWD to form an external control group is a phase 2 study to evaluate the safety and activity of blinatumomab among 189 adult patients with relapsed or refractory B-precursor acute lymphoblastic leukemia (Topp et al., 2015). The outcomes from this study were compared with those to a historical data set from Europe and the United States (Gökbuget et al., 2016). The five attributes of the primary estimand for this study are:

- *Population*: adult patients with B-precursor Ph-negative relapsed/refractory acute lymphoblastic leukemia (R/R ALL)

- *Treatment*: blinatumomab (9 ug/day for the first 7 days and 28 ug/day thereafter) by continuous intravenous infusion over 4 weeks every 6 weeks (up to five cycles) (experimental arm), or salvage therapy (possibly multiple lines) (historical control arm)

- *Endpoint*: complete remission (CR) within the first two treatment cycles in all blinatumomab treated patients (experimental arm) or after salvage therapy (historical control arm)

- *ICEs*: death before the first response assessment or adverse events leading to treatment discontinuation before the first response assessment (Topp et al., 2015)

- *Population-level summary* : Difference in CR rates for patients receiving blinatumomab and for those receiving salvage therapy

Treatment policy and while-on-treatment strategies were used as primary and secondary analyses, respectively. Note that comparability of treatment and control is always a concern when two different data sources are used and this study is no exception; importantly, the time frame for the blinatumomab trial differed substantially from that of the external control data. Several sensitivity analyses were performed using different time periods and found that there was a trend in CR rate over time when using data collected from all sites and no trend in CR rate over time when using data only from sites that provided data across the entire period (Gökbuget et al., 2016). This suggests that there may be heterogeneity of patients across sites over different time periods. Although no specific ICEs were specified for the historical control data, it is generally believed that patterns of ICEs (e.g., treatment adherence, dose changes, treatment switches or augmentation) are expected to vary between clinical trials and RWD (Sheffield et al., 2020). Burger et al. (2021) point out that the quality of external control data needs to be checked for different frequencies and/or different types of ICEs.

## 4.4  Observational studies

Patorno et al. (2019) presented an example of an observational study using RWD to mimic the CAROLINA trial (Rosenstock et al., 2019) described in Section 4.1. They identified 24,131 linagliptin and glimepiride initiators from three US claims databases (Optum Clinformatics, IBM MarketScan, and fee-for-service Medicare) for the purpose of replicating the CAROLINA trial. All estimand attributes were supposed to be the same as those in CAROLINA trial. However, some modifications were considered necessary due to the availability of RWD elements and to accommodate real-world practice.

- *Population*: patients with type 2 diabetes (T2D) at increased cardiovascular risk who initiated linagliptin or glimepiride between May 2011 and September 2015. In addition to adapting eligibility criteria from CAROLINA, the study used 1:1 PS-matching to generate comparable groups with respect to > 120 confounders

- *Treatment* : linagliptin versus glimepiride

- *Endpoint* : composite cardiovascular outcome based on the CAROLINA's primary endpoint

- *ICEs*: treatment discontinuation or switch to a comparator, nursing home admission, plan enrollment, or end of study, whichever occurred first

- *Population-level summary* : hazard ratio

The authors state that this study used an "as-treated" strategy. Patients were followed until treatment discontinuation or switch to a comparator, occurrence of event of interest, nursing home admission, plan disenrollment, or end of the study period, whichever came first. Their analytic strategy censored the subjects with ICEs; it also seems like individuals who died from causes not included in the outcome were also censored. The interpretation of the hazard ratios they report depend critically on what one assumes about censoring. If censoring is assumed to be non-informative, then their hazard ratio can refer to a world without censoring (i.e., no ICEs or death from causes not included in the outcomes). If this censoring assumption is violated, then the interpretation of hazard ratio is compromised. It is questionable whether the hazard ratios reported here are comparable to those reported in the CAROLINA trial.

## 5   A Roadmap for Choosing Appropriate Estimands

This section discusses additional considerations for defining estimands while reiterating some points made in Section 3, and presents a roadmap to choose appropriate estimands in RWE studies.

### 5.1   Who are the stakeholders and what are their research questions?

Different stakeholders often focus on different aspects of a medical product in the healthcare system and thus may ask different questions. For example:

- *Regulatory agencies* responsible for approval of new products or new indications of approved products are generally interested in the safety and efficacy or benefit-risk profile of a medical product and may ask questions regarding relative measures of primary and/or key secondary efficacy and safety endpoints between comparative groups. For example, an agency may ask: (1) What are the optimal doses of an investigational anti-diabetic product for treating adult type 2 diabetes? (2) What is the difference in response rate of a new target therapy relative to that of an SoC in treating naive patients with diffuse large B cell lymphoma? (3) What is the body-weight change at week 24 from baseline for a new therapy relative to an active control in treating adult patients with obesity?

- *Health technology assessment* (HTA) agencies may be more interested in questions of cost-effectiveness and pharmacoeconomic analysis of a new technology for a particular group of patients as compared to existing technologies (Wale et al., 2021). For example, the questions may include (1) Does the new technology work for the target patient population at a lower cost? (2) How does it compare with existing technologies in terms of cost and effectiveness? (3) What are the direct and indirect costs associated with the new technology as used in routine practice?

- *Patients* or *patient-focused organizations* such as the Patient-Centered Outcomes Research Institute (PCORI) may be interested in questions and outcomes that are "meaningful and important to patients and caregivers" in order to help them make informed decisions for their own care (Frank et al., 2014). Questions they ask may include: (1) Does the product work for me and how long do I have to take the product? (2) Will it cause any side-effects and if yes, how serious? (3) What should I do if the product does not work for me?

Even the same type of stakeholders may be interested in different aspects of a product such as optimal doses, survival time, quality of life, or benefit-risk ratio of a target therapy. Therefore, understanding who the stakeholders are and their needs helps define the research questions of interest, which in turn helps precisely defined the estimands in RWE studies.

## 5.2   What are the study objectives, designs, and analytic methods?

Although the research question drives study objectives, the same research question with similar objectives can be examined using different study designs. Different designs may require different analytic methods based on the underlying assumptions to estimate the estimands, thereby generating different levels of evidence. See also Fang et al. (2020) for some key considerations in designing RWE studies. In addition, the quality of selected RWD sources might lead to different choices for the RWE study design; see Section 5.3 for further discussion.

## 5.3   What are the fit-for-purpose RWD sources?

Availability of fit-for-purpose RWD sources is critical for RWE studies. The framework for FDA's RWE program (FDA, 2018) describes that the agency assesses fitness of RWD based on (1) reliability (data accrual and data quality control; data assurance) and (2) relevance of the selected RWD. Data reliability assessment includes data verification for completeness, consistency, plausibility, trends over time, and use of reporting standards. Data relevance assessment checks whether the data include relevant information on the disease of interest, exposure/treatment, outcomes, and covariates. It is important to ensure that RWD sources capture treatment and endpoint information that are used to define treatment effects. Data should capture sufficient information on the occurrence of pre-defined ICEs, especially those that are induced by patient behaviors and lifestyles, as well as key covariates that might be associated with treatment selection and outcome occurrence. We also refer to the ASA BIOP RWE SWG paper by Levenson et al. (2022) on statistical considerations for the evaluation of fit-for-purpose RWD for more  discussion.

## 5.4   What are the treatment regimes of interest?

TCTs usually explore the efficacy and safety profiles of a therapy compared to a pre-defined control (either a placebo or an active control). In RWE studies, however, a sequence of therapies might be of interest. The choice of therapy is influenced by many factors including physician and patient preference. For patients with a chronic disease, different lines of therapies may be needed at different stages of the disease. With an increasing number of new therapies approved in many disease areas, the number of available treatment options increases exponentially, leading to multiple possible options of treatment regimes. Therefore, articulating the primary treatment (strategy or regime) of interest in evaluating the effect is one of the key considerations in defining estimands for RWE studies, as described and emphasized in Sections 2.1 and 3.2.

## 5.5 What are the possible ICEs?

Possible ICEs in RWE studies and approaches to handling them depend on the stakeholders and their research questions. As discussed earlier, identification and specification of a precise set of ICEs have a direct impact on the definition of the estimands and subsequent estimation. The five categories of ICEs introduced in Section 3.4 may help elucidate the process of defining estimands in RWE studies. For example, an estimand can be constructed as treatment effectiveness on pre-selected endpoints (1) regardless of whether rescue therapies are used (rescue therapies are part of treatment strategies), (2) among those who can tolerate either treatment being compared (without E1 events), (3) among those who are not compliant with the current treatment due to non-behavioral factors (with E4 events, but may include those with E1, E2 and E3 events because the latter three categories may not be considered as ICEs), or (4) among patients who switch to self-preferred or physician recommended therapies. Then, the estimand in (1) is equivalent to intention-to-treat estimand. Estimands in (2) and (3) are interested in treatment effect in different principal stratum of patients. Let $E_j(a) = 1$ denote the occurrence (0 for non-occurrence) of $j$th category of ICEs under the binary treatment option $a$ defined in Section 3.4. For example, $E_1(a) = 1$ represents the occurrence of ICEs due to safety concern such as treatment intolerability under treatment option $a$, $E_2(a) = 1$ represents the occurrence of ICEs due to lack of efficacy such as treatment switch or non-adherence. Then the estimand in (2) considers a treatment effect among a set of patients say, $S_1 = \{E_1(1) = 0\} \cap \{E_1(0) = 0\}$ and can be expressed as a contrast between $Y(1)|S_1$ and $Y(0)|S_1$. Similarly, a principal stratum for the estimand in (3) can be defined as $S_4 = \{E_4(1) = 1\} \cap \{E_4(0) = 1\}$ and the estimand can be expressed as a contrast between $Y(1)|S_4$ and $Y(0)|S_4$. The decision to incorporate certain ICEs into the definition of an estimand should consider the fit-for-purpose RWD, study design, and underlying assumptions of analytic methods used to estimate the estimand. Finally, patient behaviors are extremely important in medical practice above and beyond product-induced events (e.g., AE or LoE) and thus need to be taken into consideration in defining and selecting ICEs and strategies to handle them.

## 5.6 A roadmap for choosing an appropriate estimand and RWE study design

Given what we discussed above, Figure 1 presents a roadmap to choose an appropriate estimand and associated RWE study design. Essential steps in the roadmap include (1) identification of stakeholders and their research question of interest, (2) determination of study objectives and designs, (3) finding fit-for-purpose RWD (including assessment of reliability, relevance, and benchmark assumptions for identification), (4) defining estimands based on the five attributes and corresponding estimators, (5) applying selected analytic methods with appropriate assumptions to estimate the estimands, and (6) performing sensitivity analysis to check robustness of inference to deviation of underlying assumptions.

## 6 Discussion and Conclusion

In addition to what has been discussed above, the following considerations might also be helpful (and sometimes critical) for defining estimands in RWE studies.

First, the research question may depend on how the study results will be used by stakeholders and the utility of the corresponding policy (Greifer and Stuart, 2021). More specifically, if a question concerns a treatment policy intended to apply to all qualifying patients, the target population should be the whole (indicated) patient population and estimand should be the average treatment effect (ATE). If the question concerns a policy of withholding a treatment among those currently receiving or not receiving it, the estimand should be the average treatment effect among the treated (ATT) or among the untreated. This is important to understand because different estimands require different sets of assumptions for the corresponding analysis, and interpretability of analysis results also relies on the validity of the underlying assumptions.

Second, note that choosing the fit-for-purpose RWD sources and right estimand will usually be an iterative process. The quality of RWD should be carefully assessed to ensure that clinical questions of interest can actually be answered using RWD. If the RWD source is limited, one might need to ask a modified version of the initial questions which leads to a modified version of the initial estimands. For example, if we were initially interested in ATE but all we have is product registry data, we should reconsider whether the ATE is an "addressable" question. In this case, the addressable question might be the treatment effect among those who used the product (i.e., ATT). Similarly, if a clinical question concerns the effect of expanding a therapy known to be effective to those not yet receiving it, then the fit-for-purpose RWD sources containing patients who have not received the treatment should be used to be able to define and estimate such an effect (Greifer and Stuart, 2021).

Third, although we can consider different strategies to handle a selected ICE (or a set of ICEs), interpretation of estimands based on the selected strategies should be clinically meaningful. For example, the principal stratum measuring the effect of a treatment in the subpopulation of patients for whom an ICE would not occur may not be helpful for COVID-19 trials as the interest of the trial is likely to be a treatment effect in the entire population of patients, rather than in an unknown subpopulation (Kahan et al., 2020).

In conclusion, the construction of estimands in RWE studies requires many additional considerations relative to the TCT setting. Understanding stakeholders and their questions of interest, evaluating fitness-for-purpose of RWD, and articulating each of the five attributes in E9(R1) can help identify and define appropriate estimands. Other frameworks such as causal inference or target trial frameworks can be helpful in defining RWE study estimands based on the E9(R1) framework. Analytical methods should be aligned with selected RWE study estimands and rigorous sensitivity analysis should be conducted to ensure robustness of study findings.

# 7 Acknowledgements

We would like to thank Dr. Heng Li at the FDA and Dr. Tricia Luhn at Roche for their valuable contribution to this work. Our gratitude also goes to Dr. Mark Levenson and two anonymous referees whose constructive comments and suggestions help improve the presentation of our paper.

Table 1: Similarities and differences between TCTs and RWE studies with respect to the five attributes of estimands and sensitivity analysis in E9(R1).

| Similarities/ differences | TCTs | RWE studies |
|---|---|---|
| *Population* | | |
| Similarities | • Study populations are clearly defined with a set of inclusion and exclusion criteria (IEC) | |
| Differences | • Target populations are restricted to those who meet a list of prospectively defined IEC<br>• Patients with comorbidities may be excluded<br>• Some populations such as children and elderly are usually under-represented | • IEC may be less restrictive<br>• IEC are defined based on or limited to available data<br>• Study populations may include those with comorbidities and who are under-represented in TCTs and hence generally more heterogeneous |
| *Treatment* | | |
| Similarities | • A treatment or sequence of treatments of interest is pre-specified | |
| Differences | • A single (or multiple doses of) treatment is often of interest<br>• A single sequence of treatment is often of interest<br>• One alternative treatment or multiple treatments as a whole is specified as a comparison group<br>• Use of rescue treatments may be considered as an ICE | • Various treatment use patterns are observed (non-adherence, switching, concomitant use, etc., are more frequent)<br>• Multiple treatments (some of them may be SoCs) are often involved<br>• Different lines of therapies are usually considered as more relevant (e.g., SoC followed by a new therapy) to obtain optimal dynamic treatment regimes |
| *Endpoints* | | |
| Similarities | • One or more endpoints of primary/secondary interest are defined<br>• A composite endpoint may also be used | |
| Differences | • Endpoints are usually measured at pre-defined time schedules or visits<br>• Endpoints are often measured blindly<br>• Both surrogate endpoints and clinical outcomes can be used | • Endpoints are measured only when the patient/prescriber reports the outcome<br>• Endpoints are measured without blinding<br>• Clinical outcomes are often used |

| Similarities/ differences | TCTs | RWE studies |
|---|---|---|
| *Intercurrent events (ICEs)* | | |
| Similarities | • ICEs and strategies to handle them are pre-specified | |
| Differences | • Product-induced ICEs are generally of more interest in product development<br>• Five strategies to handle ICEs as described in E9(R1) | • More diversified ICEs (than those in TCT) may occur and they can be part of treatment strategies<br>• Pattern of ICE occurrence may be more complex<br>• Patient behaviors may impact the use of a product |
| *Population-level summary* | | |
| Similarities | • Summary measures are pre-defined in study protocol<br>• The summary measures can be descriptive or comparative, followed by sensitivity analyses | |
| Differences | • Simple statistical methods are often used to estimate the estimands<br>• Results bear statistical interpretation | • Causal methods are often used to account for issues with non-randomization<br>• Results can be interpreted causally if causal assumptions hold |
| *Sensitivity analysis* | | |
| Similarities | • Analytical methods to assess robustness of results to study conclusion | |
| Differences | • Focusing on assumptions of data distribution, e.g., assumptions on statistical models<br>• Exploration of statistical bias and different missing data mechanisms<br>• Often analyses of multiple subsets (e.g., subgroups), multiple outcomes, and multiple imputation methods performed | • Focusing on causal assumptions, e.g., consistency, positivity, and exchangeability<br>• Decomposition of statistical bias and identification bias<br>• Analyses based on one or more causal and/or statistical assumptions simultaneously |

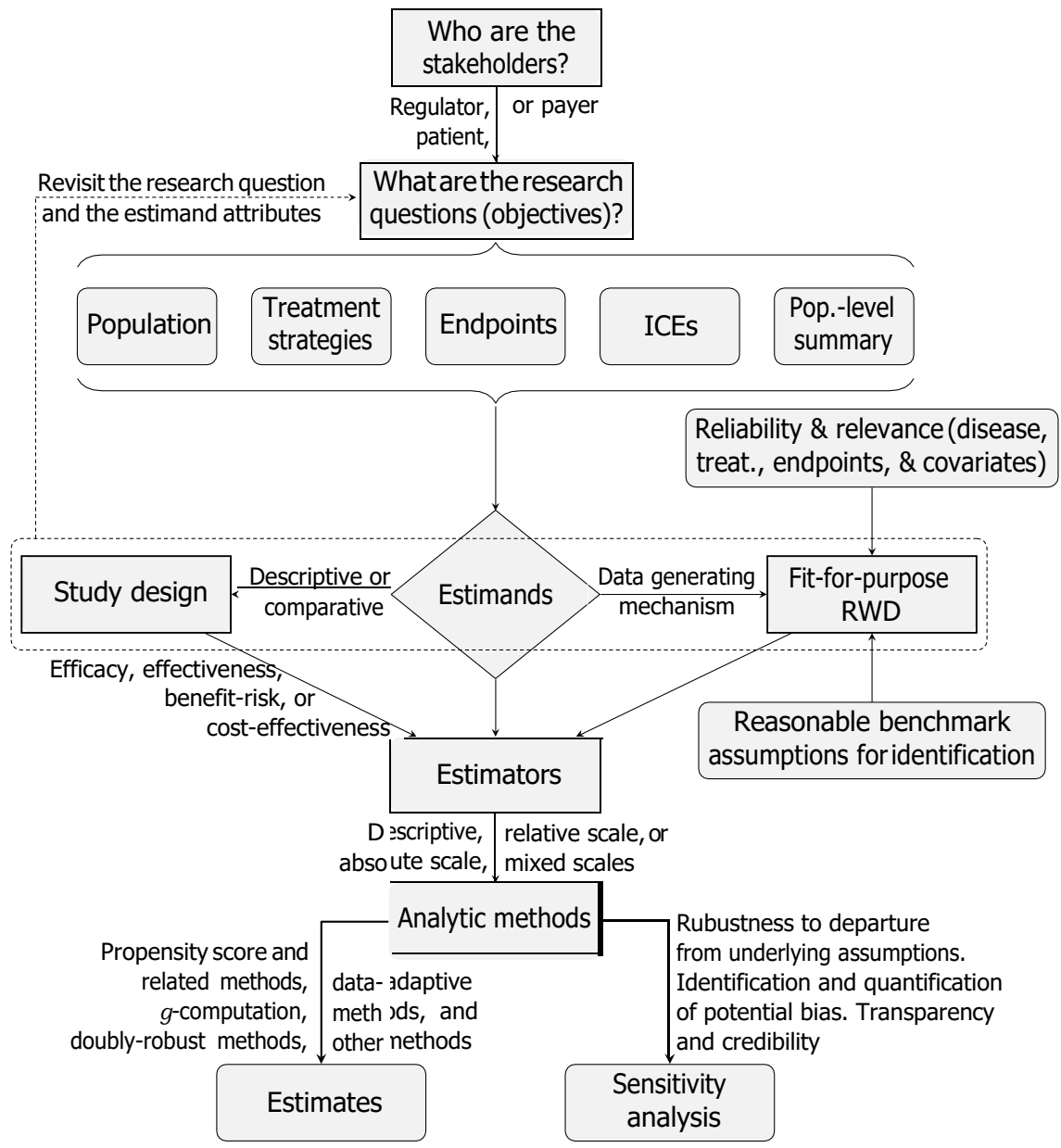

Figure 1: A roadmap for formulation of an RWE study, starting from stakeholders with specific research questions, choosing appropriate estimands by considering their five attributes, selecting the fit-for-purpose RWD (including assessment of reliability, relevance, and benchmark assumptions for identification), designing a study, applying selected analytic methods with appropriate assumptions to estimate the estimands, and performing sensitivity analysis to evaluate robustness of study findings.